\documentclass[twocolumn,preprintnumbers,amsmath,amssymb,prd]{revtex4}
\usepackage{hyperref}
\usepackage{graphicx} % Include figure files
\usepackage{dcolumn}  % Align table columns on decimal point
\usepackage{bm}       % bold math
\usepackage{mathrsfs} %script fonts

\def\half{\tfrac{1}{2}}           % small built-up `one-half'
\newcommand{\scri}{\mathscr{I}}   % Scri
\newcommand{\Lie}{\mathfrak{L}}   % Lie derivative

\begin{document}

\preprint{gr-qc/0512167}
\preprint{UM-PP-05-54}

\title{Over the Rainbow: Numerical Relativity beyond Scri+}

\author{Charles W.\ Misner}
\email{misner@physics.umd.edu}
\affiliation{Department of Physics, University of Maryland,
College Park, MD 20742-4111, USA}
\affiliation{
Albert-Einstein-Institut, Max-Planck-Institut 
f\"ur Gravitationsphysik, Am M\"uhlenberg 1, D-14476 Potsdam, Germany
}

\date{12 March 2006}
%% 20 Jan 2006 added Tiglio references: v2      %%%%%%%%%%%%
%% 12 Mar 2006 added acknowledgement section         %%%%%%%
%%             and reference to 'das All' eprint: v3 %%%%%%%

\begin{abstract}

This is a study of the behavior of wave equations in conformally
compactified spacetimes suited to the use of computational boundaries
beyond Scri+. 
There light cones may be adjusted for computational convenience and/or
Scri+ may be approximated by a ``proto-Scri" spacelike hypersurface just
outside a de~Sitter horizon. One expects a numerical implementation to
excise the physically unnecessary universe somewhat beyond the outer
horizon. 
As an entry level example I study forms of the Maxwell equations
and causal relations for an outer boundary in that example. 
\end{abstract}

%\pacs{Valid PACS appear here}% PACS, the Physics and Astronomy
                             % Classification Scheme.
%\keywords{Suggested keywords}%Use showkeys class option if keyword
                              %display desired
\maketitle

\section{\label{sec:Intro}Introduction}

The focus of this essay is the matching of a computational model of a
source of gravitational waves (such as a compact object binary) to the
cosmological environment through which any emitted signal travels to a
detector. 
The transmission phase is well understood and is essentially the
propagation of a linearlized gravitational wave through a background
cosmology which includes such effects as a cosmological redshift and
gravitational lensing due to intervening masses, plus the kinetic
effects of the detector's motions.
For these purposes the geometrical optics approximation is often
adequate. 
Although implementing this transmission model in observatory data
analysis demands skill, it is simpler than the electromagnetic case
where interactions with dust, gas and plasmas en route must also be
considered.
In this large scale background the source can be taken to be a point
source of linearized gravitational waves having some specific antenna
patterns and wave forms.
The more unsettled question then is how to extract from a detailed model
of a wave generation system the appropriate linearized wave.
The most common assumption is that the wave generator can be modeled 
as existing in an asymptotically flat spacetime where waves propagating 
out to future null infinity (Scri+ or $\scri^+$) will be described 
there by linearlized gravitation and inserted as the point source 
is the cosmological context.
This is clearly justified by the huge difference in scales between 
the size of the wave generator and the radius of curvature of its 
surrounding spacetime (the cosmos or a galaxy or stellar cluster).
The question then is how to efficiently calculate and extract the
waveforms at $\scri^+$.

In the usual Cauchy evolutions on asymptotically flat spacelike
hypersurfaces, the waves must be recognized at large distances from the
source, possibly a long time after the source motions have ceased
violent activity. 
One way to overcome this time lag is through the use of a retarded time
coordinate. This works well in spherical symmetry \cite{HM:retd,
BST:retd}, but is poorly defined otherwise. 
The Pittsburgh group has used such characteristic evolution at larger
distances while matching to more common Cauchy evolution in the interior
\cite{win:lrr}.
The use of hyperboloidal time slices (some generalization of
hyperboloids in Minkowski spacetime) appears attractive as a smooth
transition from conventional time slices in central regions to
asympotically null spacelike hypersurfaces toward infinity.
Were such a slicing to be combined with a straightforward Cauchy
evolution, the Courant time step condition would defeat the advantages
of approximating retarded time, as the nearly infinite speed of light
toward infinity would require infinitesimal time steps.
Thus hyperboloidal slicings effectively require a conformal
compactification which makes the coordinate speed of light finite again
\cite{MSL:KITP}.

While many other approaches \cite{hf:sh,sascha:tuebingen,win:lrr}
reformulate the Einstein equations, I conjecture that the ordinary
Einstein equations might be used for a conformally compactified metric
$g_{\mu\nu} = \Omega^2 \tilde{g}_{\mu\nu}$ where $g_{\mu\nu}$ is the
compactified metric and $\tilde{g}_{\mu\nu}$ is the physical metric. 
If the compactification is successful both metrics will have nonsingular
Einstein tensors where $\Omega = 0$; the physical metric because it has
a vanishing Einstein tensor, and the compactified metric because the
null hypersurface where $\Omega = 0$ is by assumption smoothly imbedded
in the conformally regulated spacetime with metric $g_{\mu\nu}$. 
Thus the Einstein equations should reduce to the Einstein tensor of
$g_{\mu\nu}$ set equal to an effective stress energy tensor computed
from the conformal factor $\Omega$ as described in \cite[Equation 
11.1.16]{wald:wald}.
Although these terms contain $\Omega$ in denominators, this effective
stress-energy tensor must be regular since the conformally regulated
metric and its Einstein tensor are. 
Before studying this conjecture in constructive detail, I want to
confirm that wave equations, evolved numerically by conventional Cauchy
evolution schemes, do not fail for some unanticipated cause. 
Thus this paper formulates a problem for numerical study in which the
Maxwell equations are to be evolved in a conformally compactified
spacetime that is nearly Minkowskian.

In addition to the use of hyperboloidal time slices and conformal
compactification, I suggest two additional tools to ameliorate the
extraction of outgoing waves in numerical evolutions. 
One is the use of artificial cosmology to convert the null hypersurface
$\scri^+$ into a spacelike hypersurface in the future of the initial
Cauchy hypersurface as proposed in \cite{CWM:DeserF}. 
This procedure fattens $\scri^+$ into a shell of spacelike hypersurfaces
by changing the value of the cosmological constant plausible for the
present universe by many orders of magnitude so that the de~Sitter
horizon is located, not at cosmological distances, but merely well
beyond the region where the wave generating mechanism can be influenced
by it. 
In this sense it is similar to the computational tool of artificial
viscosity which fattens a shock from its physical thickness on the
order of the molecular mean free path into a much larger length which is
still small in the range of scales that otherwise are met in the
physical problem at hand. 
The second additional tool is the tilting of the null cones governing
wave propagation in ways that ease the computational behavior in the
unphysical (`Oz') region beyond the de~Sitter horizon and beyond
$\scri^+$.

\section{\label{sec:deSitterMetric}Background Spacetime}
As in \cite{CWM:DeserF} the background metric for this study will
be the de~Sitter spacetime
\begin{equation} 
ds^2 = -dT^2 +dX^2 + dY^2 + dZ^2 + (R^2/L^2)(dT-dR)^2  
\label{eq:deSorig} 
\end{equation} 
with an artificial cosmological constant $\Lambda = +3/L^2$. We are
interested in the limit $1/L^2 = 0$ when this becomes Minkowski
spacetime, and the cosmological constant is used to make boundary
conditions numerically simpler at the de~Sitter horizon than
they would be at flat spacetime's $\scri^+$. 
The de~Sitter horizon is the null hypersurface $R=L$ in this metric. 

As before I introduce the coordinate changes 
\begin{subequations}
\label{eq:Hslice} 
\begin{equation} \frac{T}{s} = u +
    \frac{\tfrac{1}{2}r^2}{1-\tfrac{1}{4}r^2} \quad . 
    \label{eq:HsliceU}
\end{equation} 
\begin{equation} 
    \frac{X^i}{s} = \frac{x^i}{1-\tfrac{1}{4} r^2} \quad . 
    \label{eq:AnMR}  
\end{equation}
\end{subequations} 
where $R^2 = X^2 +Y^2 +Z^2 \equiv X^i X^i$ and $r^2 = x^2+y^2+z^2 \equiv
x^i x^i$. 
This brings $\scri^+$ in to $r=2$ in the Minkowski case and in the de
Sitter case makes this a spacelike hypersurface beyond the de~Sitter
horizon which I call proto-$\scri^+$.
The hypersurfaces of constant $u$ are then hyperboloids 
\begin{equation}
    \label{eq:hyperboloid} 
    [T-s(u-1)]^2 - R^2 = s^2 
\end{equation} 
in Minkowski spacetime and are also asymptotically null spacelike
hypersurfaces in the de~Sitter modification. 
This leads to a metric which is singular at $r=2$, but only in a
conformal factor $s^2/q^2=s^2/(1 - \tfrac{1}{4} r^2)^2$ which does not
appear in the Maxwell equations (see \cite{CWM:DeserF}). 
Thus we can drop this conformal factor and our test problem is to solve
the Maxwell equations in the resultant metric which is of the usual 3+1
form
\begin{equation}
       ds^2 = - \alpha^2 dt^2 
              + \gamma_{ij}(dx^i + \beta^i dt) (dx^j + \beta^j dt)
\label{eq:g3+1}
\end{equation}
with $t=u$,
\begin{subequations} 
     \label{eq:CRdeSitter} 
\begin{equation} 
      \alpha^2 = {\alpha_0}^2 \left[ 1+\frac{s^2}{L^2} S(r)^2 \right]^{-1} \quad ,
      \label{eq:alpha} 
\end{equation} 
\begin{equation}  
      \beta^k = (x^k /r)\beta
      \quad , 
\end{equation} 
\begin{equation}  
      \beta = -r \left[1+\frac{s^2}{L^2} S(r) \right] 
               \left[1+\frac{s^2}{L^2 }S(r)^2 \right]^{-1}
      \label{eq:beta} 
\end{equation} 
\text{and} 
\begin{equation} 
      \gamma_{ij} = \delta_{ij} + (s/L)^2 \frac{x_i x_j}{r^2} S(r)^2 \quad .
      \label{eq:gamma} 
\end{equation} 
\end{subequations} 
The definitions 
\begin{subequations}
\label{eq:Sa0}
\begin{equation}
\label{eq:S}
     S(r)  =  \frac{r}{(1+\half r)^2} 
\end{equation}
\text{and}
\begin{equation}
       \alpha_0  =   1+ \tfrac{1}{4} r^2 \quad     
\label{eq:a0}
\end{equation}
\end{subequations}
yield the regulated de~Sitter metric, but could be modified outside $r=2$
to control error propagation in the unphysical region.

To see the behavior of the light cones for this regulated metric, we
calculate the coordinate speed of light $dr/du$ in radial null
directions.
This gives
\begin{equation}
\label{eq:lightspeed}
  v = \frac{r + r (s^2/L^2) S \pm \alpha_0}{1 + (s^2/L^2) S^2}
    \quad .
\end{equation}
For the de~Sitter values above this reduces to 
\begin{equation}
\label{eq:vout}
   v_{\rm out} = (1+ \half r)^2
\end{equation}
and 
\begin{equation}
\label{eq:vin}
  v_{\rm in} = \frac{-(1-\half r)^2 + (s^2/L^2)(1+\half r)^2 S^2}%
    {1 + (s^2/L^2) S^2}
  \quad .
\end{equation}
Note that the constant $r$ hypersurfaces are never spacelike
when $s/L=0$ since then the inward speed of light is $v_{\rm in} \leq 0$.
When the de~Sitter option is used one finds $v_{\rm in} > 0$, i.e., both
the inward and the outward sides of the lightcone point toward
increasing $r$, for an interval around $r=2$.

\section{Tilt}

In evolution codes which permit setting the computational boundary on a
coordinate sphere (such as those described in
\cite{kls05:boundary,thornburg:boundary,lsu:lrt,lsu:ddst}) the outer
boundary could be set at $r=2$ or slightly beyond in the metric
described above (or ones asymptotically similar).
However, if a cubic boundary enclosing $\scri^+$ is to be used, there
are additional behaviors beyond $\scri^+$ that need attention.
For example, if the artificial de~Sitter horizon is set at $L = 10 s$, 
then the $r=\text{constant}$ hypersurfaces are spacelike only in the 
narrow interval of (approximately) $r=1.81$ to $r=2.21$.
But a boundary cube enclosing $r=2$ has corners at $r=2 \sqrt{3} \approx 
3.46$.
Thus on most of such a cubic boundary the light cones (from equation 
\ref{eq:vin}) would allow inwardly propagating waves, requiring careful 
attention to the boundary conditions which one hopes to avoid.
In this case I suggest that the metric be modified outside $r=2$ so that
all $r=\text{constant}$ hypersurfaces beyond $r=2$ are spacelike ---
i.e., so that $v_{\rm in} > 0$ in the entire computational domain beyond
$r=2$.
Analytically this would have no effect on the physical solution inside
$r=2$ provided all fields propagate causally.
It is also straightforward to do in the test case of a Maxwell field
with a fixed background metric.
If these methods were to be developed for the Einstein equations
(without using a spherical boundary such as $r=2$) it would necessitate
distinguishing, in the Einstein equations beyond $r=2$, the $g_{\mu\nu}$
metric begin evolved from the appearances of the metric as coefficients
of the principle derivatives which determine the causal relations in the
evolution algorithm.

In the Minkowski case $s/L=0$ (no artificial cosmology) it is not
possible to make a $C^2$ change in the metric only beyond $r=2$ which
avoids inward pointing null rays just beyond $r=2$ --- in that case one
has, from the $r \leq 2$ formula above for $v_{\rm in}$ that
$(d^2/dr^2)v_{\rm in}= -\half$ at $r=2$ so that $v_{\rm in}$ must have a local
maximum value of zero at that point and is necessarily negative (inward)
in a neighborhood of that point, including some $r>2$ points. 
For the de~Sitter case there is again a maximum of $v_{\rm in}$ near $r=2$,
but its value there is positive so it can be smoothly turned upward (to
avoid zero or negative values) in suitable intervals beyond the
de~Sitter horizon (and beyond $r=2$ for small values of $s/L$).

\section{\label{sec:MaxEq0}Maxwell's Equations with Conserved Constraints}

We take the Maxwell equations in the form given in the DeserFest paper
\cite{CWM:DeserF}.
The quantities involved are all various components of the usual
4-dimensional Maxwell fields $F_{\mu\nu}$ and $\mathfrak{F}^{\mu\nu} =
\sqrt{-g}g^{\mu\alpha}g^{\nu\beta} F_{\alpha\beta}$.
In particular we define,
$     \mathfrak{F}^{0i} = \mathfrak{D}^i
$
, 
$     \mathfrak{F}^{ij} = \mathfrak{H}^{ij} = [ijk] H_k
$
and 
$
F_{i0} = E_i$, $F_{ij} = B_{ij} = [ijk] \mathfrak{B}^k
$.
Here $[ijk]$ is the completely antisymmetric symbol $[ijk] = 0, \pm
1$ with $[123] = +1$.  
The index $0$ here refers to the coordinate basis vectors $\partial_t$
or $dt$.
The fundamental fields in the formulation are $\mathfrak{B}^i$
and $\mathfrak{D}^i$. The constraints read
\begin{equation}
  \partial_i \mathfrak{B}^i = 0 = \partial_i \mathfrak{D}^i 
  \quad .
  \label{eq:constr}
\end{equation}
In terms of these primary fields, two auxiliary fields ${E}_i$ and
${H}_i$ are computed by the formulae 
\begin{subequations}
\label{eq:defEH}
\begin{equation}
\label{eq:defE}
  {E}_i = (\alpha/\sqrt{\gamma})\gamma_{ij}\mathfrak{D}^j
    + [ijk]\beta^j \mathfrak{B}^k 
\end{equation}
and
\begin{equation}
\label{eq:defH}
  {H}_i = (\alpha/\sqrt{\gamma})\gamma_{ij}\mathfrak{B}^j 
     - [ijk]\beta^j \mathfrak{D}^k
      \quad .
\end{equation}
\end{subequations}
The evolution equations are then
\begin{subequations}
\label{eq:BDevol}
\begin{equation}
\label{eq:Bevol}
   \partial_t \mathfrak{B}^i  = - [ijk]\partial_j E_k
\end{equation}
and
\begin{equation}
\label{eq:Eevol}
   \partial_t \mathfrak{D}^i = [ijk]\partial_j H_k
   \quad .
\end{equation}
\end{subequations}
Because the combinations of metric components appearing in equations 
(\ref{eq:defEH}) are each conformally invariant, it follows that field
components $\mathfrak{B}^i$, $\mathfrak{D}^i$, $E_i$, and $H_i$ 
satisfying these Maxwell equations are also each conformally invariant.

By differentiating equations (\ref{eq:BDevol}) one find equations 
for the evolution of the constraints
\begin{equation}
\label{eq:cevol0}
   \partial_t (\partial_i \mathfrak{D}^i) = 0 
        = \partial_t (\partial_i \mathfrak{B}^i) \quad .
\end{equation}
These show that the constraints are conserved, so that if they are 
initially zero they remain zero within the limits of the numerical 
approximations used.
This sort of behavior has been found less than optimal as shown, e.g., in 
a Maxwell example by Baumgarte et al.\ \cite{baumgarte:em}.  An alternative 
form for the equations is given in section \ref{sec:MaxEqN} below 
which emphasizes use of the direction normal to the time slices.

\section{\label{sec:Energy}Energy Conservation}

Because the metrics here are stationary with Killing vectors $\xi =
\partial_0 = \partial_t$ there is a conserved energy. The covariant
``conservation'' law ${T^{\mu\nu}}_{;\nu}=0$ leads to a true conservation
law $\partial_\nu \rho^\nu = 0$ where $\rho^\nu = -\xi^\mu \sqrt{-g}
{T_\mu}^\nu$ which gives a conserved integral $\int \rho^t\,d^3x$ whose
changes can be monitored by the energy flow through a boundary integral.
Most of these equations hold only in a coordinate basis, so we express
the energy current components using the coordinate basis components 
of the Maxwell field from section \ref{sec:MaxEq0} above.

For the electromagnetic field the stress-energy tensor density is
given by
\begin{equation}
  \label{eq:maxTmunu}
   4\pi \sqrt{-g}{T_\mu}^{\nu} = F_{\mu\alpha}\mathfrak{F}^{\nu\alpha}
        -\tfrac{1}{4} \delta_\mu^\nu  F_{\alpha\beta}\mathfrak{F}^{\alpha\beta}
     \quad .
\end{equation}
When the components in 3+1 form  are inserted with $\xi^\mu
= (1;0,0,0)$ one finds for the energy density
\begin{equation}
  \label{eq:EnDensity}   
     4\pi \rho^t = \half(\mathfrak{D}^k E_k + \mathfrak{B}^k H_k)
\end{equation}
and for the energy flux
\begin{equation}
  \label{eq:EnFlux}
     4\pi \rho^i = [ijk] E_j H_k
     \quad .
\end{equation}
These expressions, in addition to the constraints, provide possible
checks to monitor the validity of numerical implementations of the
Maxwell field evolution. 
Note that these expressions are conformally invariant (since each field
is) and yield spatial coordinate independent integrals as no further
metric factors enter when forming integrals of these densities.

\section{\label{sec:MaxEqN}Maxwell's Equations with Dispersing Constraints}

I here reformulate the Maxwell equations from the DeserFest paper
\cite{CWM:DeserF} and section \ref{sec:MaxEq0} above to make the
principal terms invoke the derivative along the normal $\partial_N =
\partial_t - \beta^k \partial_k $. 
This makes it natural to work with components of the fields taken in a
frame whose time axis is normal to the $t=\text{const}$ hypersurfaces. 
Thus I use the basis
\begin{subequations}
\label{eq:normal}
\begin{eqnarray}
\label{eq:enorm}
     e_N & = & \partial_N = \partial_t - \beta^k \partial_k \quad ,
   \\
     e_i & = & \partial_i  \quad ,
        \nonumber
\end{eqnarray}
\text{and dual basis}
\begin{eqnarray}
       \omega^N & = &  dt \quad ,
             \\
       \omega^k & = & dx^k + \beta^k dt
     \nonumber    
\label{eq:wnormal}
       \quad .
\end{eqnarray}
\end{subequations}
Note that this basis is constructed in a conformally invariant way since
the shift $\beta^i$ is a conformally invariant aspect of the metric
tensor.
Tensor components will be identified by a tilde when referred to this
basis. 
%% (Although a tilde on the index might be more logical, it would
%% be less visible.) 
Thus the metric components $\tilde{g}_{\mu\nu}$ can be read from
equation (\ref{eq:g3+1}) and are $\tilde{g}_{NN} = -\alpha^2$,
$\tilde{g}_{Ni} = 0$, and $\tilde{g}_{ij} = \gamma_{ij} = g_{ij}$. 
For the other fields of interest we define
\begin{subequations} 
 \label{eq:tilde}
\begin{equation}
  A = - \psi dt + A_i dx^i = -\tilde{\psi} \omega^N + \tilde{A}_i \omega^i
     \quad ,
\end{equation}
\begin{equation}
  F = -\tilde{E}_i \omega^N \wedge \omega^i 
           + \half \tilde{B}_{ij} \omega^i \wedge \omega^j 
      \quad ,
\end{equation}
\text{and}
\begin{equation}
  \mathfrak{F} = \tilde{\mathfrak{D}}^i e_N \wedge e_i 
            + \half \tilde{\mathfrak{H}}^{ij} e_i \wedge e_j
    \quad .
\end{equation}
\end{subequations}
By inserting the basis vectors from equations (\ref{eq:normal}) these
can be expressed in terms of the coordinate based components used in
\cite{CWM:DeserF} and section \ref{sec:MaxEq0} and one finds that the
principle fields are the same:
$\tilde{A}_i = A_i$, $\tilde{B}_{ij} = B_{ij}$, and
$\tilde{\mathfrak{D}}^i = \mathfrak{D}^i$. 
The subsidiary field components do change
\begin{subequations} 
   \label{eq:tilde2coord}
\begin{equation}
   \psi =  \tilde{\psi} - \beta^k A_k
     \quad ,
\end{equation}
\begin{equation}
  E_i = \tilde{E}_i +B_{ij}\beta^j = \tilde{E}_i + [ijk]\beta^j \mathfrak{B}^k
                 \quad ,
\end{equation}
\text{and}
\begin{equation}
  \mathfrak{H}^{jk} = \tilde{\mathfrak{H}}^{jk} 
          -\beta^j \mathfrak{D}^k + \beta^k \mathfrak{D}^j
\end{equation}
\text{or}
\begin{equation}
  H_i = \tilde{H}_i - [ijk]\beta^j \mathfrak{D}^k
                 \quad .
\end{equation}
\end{subequations}
The antisymmetric 3-tensors have been renumbered in a metric independent
way to 3-vectors of a different density by relations such as 
\begin{equation}
\label{eq:newH}
       H_k  = \half [kij] \mathfrak{H}^{ij}  
  \quad \text{or} \quad
       \mathfrak{H}^{ij} =  [ijk] H_k 
\end{equation}
and similarly $\mathfrak{B}^k  = \half [kij] B_{ij}$.

As the fundamental fields being evolved remain $\mathfrak{D}^i$ and
$B_{ij}$ or $\mathfrak{B}^k$, the constraints are the unchanged
equations (\ref{eq:constr}) above.
Evolution equations in terms of the frame based components are found by
substituting equations (\ref{eq:tilde2coord}) into the evolution
equations (\ref{eq:BDevol}). 

In the slicing normal frame (\ref{eq:normal}) it is straightforward to
relate the $\mathfrak{F}$ components to those of $F$ since there are no
time-space metric components $\tilde{g}_{Ni}$ in that basis. Thus we
find
\begin{equation}
\label{eq:defH2}
   B_{ij} = (1 / \alpha \sqrt{\gamma}) 
                  \gamma_{ik}\gamma_{jl}\tilde{\mathfrak{H}}^{kl}
\end{equation}
or  
\begin{subequations}
\label{eq:defEH1}
\begin{equation}
\label{eq:defH1}
  \tilde{H}_i = (\alpha/\sqrt{\gamma})\gamma_{ij}\mathfrak{B}^j
      \quad .
\end{equation}
\text{and}
\begin{equation}
\label{eq:defE1}
  \tilde{E}_i = (\alpha/\sqrt{\gamma})\gamma_{ij}\mathfrak{D}^j
\end{equation}
\end{subequations}
To obtain equation (\ref{eq:defH1}) from (\ref{eq:defH2}) does, however,
need the definition of a determinant in the form 
\begin{equation}
\label{eq:det}
  [ijk] \gamma_{ip} \gamma_{jq} \gamma_{kr} = [pqr] \gamma
      \quad .
\end{equation}

The evolution equations for $\mathfrak{B}$ and $\mathfrak{D}$ 
found in this way are 
\begin{subequations}
\label{eq:evol}
\begin{equation}
\label{eq:Bevol1}
       (\partial_t - \Lie_\beta) \mathfrak{B}^i  
            =  -[ijk]\partial_j \tilde{E}_k - \beta^i (\partial_k \mathfrak{B}^k)
\quad
\end{equation}
\text{and}
\begin{equation}
\label{eq:Devol1}
   (\partial_t - \Lie_\beta) \mathfrak{D}^i = 
         [ijk]\partial_j \tilde{H}_k  -  \beta^i (\partial_k \mathfrak{D}^k)
\quad .   
\end{equation}
\end{subequations}
In these equations the Lie derivative can be expanded as
\begin{equation} 
    \label{eq:Lie}
    \Lie_\beta \mathfrak{B}^i = \partial_j (\beta^j \mathfrak{B}^i) 
                - \mathfrak{B}^j \partial_j\beta^i
\end{equation}
and similarly for $\Lie_\beta \mathfrak{D}^i$.
These equation (\ref{eq:evol}) are unchanged, just rearranged, from
(\ref{eq:BDevol}). 
But on the right hand side of each of equations (\ref{eq:evol}) the
terms containing $\beta$ vanish in view of the constraints. 
Thus these terms can be omitted, leading to 
\begin{subequations}
\label{eq:evol2}
\begin{equation}
\label{eq:Bevol2}
       (\partial_t - \Lie_\beta) \mathfrak{B}^i  
            =  -[ijk]\partial_j \tilde{E}_k
\end{equation}
\text{and}
\begin{equation}
\label{eq:Devol2}
   (\partial_t - \Lie_\beta) \mathfrak{D}^i = 
         [ijk]\partial_j \tilde{H}_k 
                \quad .   
\end{equation}
\end{subequations}
This system (\ref{eq:evol2}) is physically equivalent to (\ref{eq:BDevol})
--- i.e., agrees with them when the constraints are satisfied --- but
may have different behavior in numerical approximations when the
constraints will not be exactly zero. 
This is in fact the case, as the evolution equations (\ref{eq:BDevol})
leave the constraint values constant in time, while the evolutions
system (\ref{eq:evol2}) show a different behavior as I now show. 

By taking the divergence of equations (\ref{eq:evol2}) one finds
\begin{subequations}
\label{eq:cevol2}
\begin{equation}
\label{eq:divBevol2}
       (\partial_t - \beta^i \partial_i) (\partial_k \mathfrak{B}^k)  
            =  (\partial_i \beta^i) (\partial_k \mathfrak{B}^k)
                \quad 
\end{equation}
\text{and}
\begin{equation}
\label{eq:divDevol2}
       (\partial_t - \beta^i \partial_i) (\partial_k \mathfrak{D}^k)  
            =  (\partial_i \beta^i) (\partial_k \mathfrak{D}^k)
                \quad .   
\end{equation}
\end{subequations}
These are each ordinary differential equations for the respective
constraints along curves with tangents $(\partial_t - \beta^i
\partial_i)$ along which the constraints will have an exponential
behavior with e-folding time $(\partial_i \beta^i)^{-1}$.
Although all the equations in sections \ref{sec:MaxEq0},
\ref{sec:Energy}, and \ref{sec:MaxEqN} are valid for any metric of the
form (\ref{eq:g3+1}), we are most interested in the regulated de~Sitter
metric defined by equations (\ref{eq:CRdeSitter}) through
(\ref{eq:Sa0}).
There one calculates that (for small values of $s/L$) to good
approximation $\beta^i = -x^i$ and $\partial_k \beta^k = -3$ so that the
constraint propagation equations (\ref{eq:cevol2}) show the constraints
propagating along the normals to the constant $u$ hypersurfaces while
decaying exponentially with an e-folding time of $\Delta u = 1/3$.
It may not be possible to preserve this favorable constraint behavior if
the metric is modified by tilting the light cones beyond $\scri^+$.

\begin{acknowledgments}
I thank colleagues at AEI Potsdam for support and expert advice as this
work was being done, and I thank James van Meter and David Fiske in the
Maryland area for motivating me to complete this paper by offering to
test some of its ideas numerically \cite{VFM:CRMaxwell}.
\end{acknowledgments}

\bibliography{Misner}

\end{document}